# 2    FUTURE E+/E- RING COLLIDERS

## 2.1    DAΦNE Consolidation Program and Operation with the KLOE-2 Detector

Catia Milardi, David Alesini, Maria Enrica Biagini, Simone Bini, Manuela Boscolo, Bruno Buonomo, Sergio Cantarella, Antonio De Santis, Giampiero Di Pirro, Giovanni Delle Monache, Alessandro Drago, Luca Foggetta, Oscar Frasciello, Alessandro Gallo, Riccardo Gargana, Andrea Ghigo, Francesco Guatieri, Susanna Guiducci, Franco Iungo, Carlo Ligi, Andrea Michelotti, Luigi Pellegrino, Ruggero Ricci, Ugo Rotundo, Giancarlo Sensolini, Angelo Stella, Alessandro Stecchi, Mikhail Zobov
Mail to: catia.milardi@lnf.infn.it
LNF-INFN, Via E. Fermi, 40 I-00044 Frascati (Rome), Italy;
Dmitry Shatilov
BINP SB RAS, Novosibirsk, Russia;
Alexander, Valishev
Fermilab, Batavia, USA.

### 2.1.1    Introduction

After a long preparatory phase, including a wide hardware consolidation program, the Italian lepton collider DAΦNE, is now systematically delivering data to the KLOE-2 experiment.

In approximately 200 days of operation 1 fb$^{-1}$ has been given to the detector limiting the background to a level compatible with an efficient data acquisition.

Instantaneous and maximum daily integrated luminosity measured, so far, are considerably higher with respect to the previous KLOE runs, and are: $L_{ist} \sim 2.0 \cdot 10^{32}$ cm$^{-2}$s$^{-1}$, and $L_{\int day} \sim 12.5$ pb$^{-1}$ respectively.

A general review concerning refurbishing activities, machine optimization efforts and data taking performances is presented and discussed.

The DAΦNE [1] accelerator complex consists of a double ring lepton collider working at the c.m. energy of the Φ-resonance (1.02 GeV) and an injection system. The collider is based on two independent rings, each ~97 m long, sharing an interaction region, where a detector is installed. The full energy injection system including an S-band linac, 180 m long transfer lines and an accumulator/damping ring provides electron–positron injection in topping-up mode during luminosity delivering.

Long radiation damping times, low collision energy and high stored currents make achieving high luminosity at DAΦNE a quite challenging task. In fact, best performances in terms of luminosity have been attained only after several radical modifications [2,3], with respect to the original design, and after implementing the new *Crab-Waist* collision scheme [4,5,6].

The highest instantaneous luminosity, $L = 4.5 \cdot 10^{32}$ cm$^{-2}$s$^{-1}$, has been measured at DAΦNE during a test run with a table-top experimental apparatus without solenoidal field. Such luminosity, two orders of magnitude higher than the best ever achieved at other colliders working at the same c.m. energy, opened new perspectives for the KLOE experiment [7]. Integrating the high luminosity collision scheme with a large detector, having a strongly perturbing solenoidal field, posed new challenging issues concerning layout, beam acceptance and coupling correction. A new interaction region has been designed, and has been equipped with a transverse betatron correction mechanism, based on rotated quadrupoles and anti-solenoids, independent for the two beams [8].

Operations for the KLOE detector have been organized in two stages. First the new interaction region has been installed, tested and the luminosity, already measured in the pre-*CrabWaist* configuration, has been reproduced and slightly improved, L = 1.52•$10^{32}$ cm$^{-2}$s$^{-1}$ [9]. Then, in the first seven months of 2013, the accelerator complex has been shut down again mainly to upgrade the detector, KLOE-2, which in view of a higher luminosity, had extended its physics search program, and to implement a general consolidation program concerning the machine hardware.

The KLOE-2 setup includes new tracking and calorimeter devices close to the interaction region. A very light tracker, consisting of a cylindrical gas electron multiplier, has been installed in the tight space between the drift chamber and the spherical beam pipe. Crystal calorimeters have been inserted in front of the collider low-β quadrupoles, thus increasing the acceptance for photons emitted under a very low angle, a key issue for rare decay studies. Dedicated detectors have been inserted one inside the experimental apparatus and the other after the first dipole in the long arc in each ring in order to study scattered electron and positron produced in γ-γ reactions.

Such upgrade imposed the extraction of the Interaction Region from inside the detector and the disassembly of the low-β section. As a consequence the collider had to be commissioned again nearly from scratch. After achieving reasonable performances, an eight months period has been dedicated to the optimization of the experiment data taking. In this phase all the activities have been addressed to demonstrate that DAΦNE was able to provide high rate, high quality physics events to KLOE-2 experiment, in a stable and reproducible way over the long term.

## 2.1.2 DAΦNE Consolidation Program

When KLOE was reinstalled on the collider Interaction Region (IR), at the end of 2010, the DAΦNE infrastructure had been working since more than 17 years.

Commissioning, aimed at setting up the collider, was affected by many, relevant and time consuming faults, and pointed out some shortcomings in the interaction region mechanical design.

In fact, several sub-systems, relying on obsolete technologies, suffered from spare part shortage. Some components got seriously damaged. It is the case of some bellows in the IR, which had lost electrical continuity causing anomalous beam induced heating of one of the two defocusing quadrupoles, resulting in a harmful random vertical beam tune-shift. The mechanical structure of the Interaction Region (IR) had shown to be inadequate to steadily support the heavy defocusing quadrupoles cantilevered inside the detector. As a consequence the two beams were oscillating in phase at 10 Hz in the vertical plane.

In this context, the shutdown scheduled in 2013, intended mainly to install new detector layers inside the KLOE detector (KLOE-2) [10], offered a very convenient opportunity to undertake a wide consolidation program involving several subsystems as well as to revise the IR mechanical design [11].

Many other machine improvements have been implemented during the following operation periods, profiting from season and maintenance shutdowns, and the unexpected faults, whose occurrence is, anyhow, diminishing in number and importance.

### 2.1.2.1 *Mechanical Upgrade*

A major effort has been done to upgrade the Interaction Region mechanical structure and the vacuum chamber around the Interaction Point.

The vacuum chamber around the Interaction Point (IP) has been replaced. The new one has tapered transition between the thin ALBEMET sphere and the Al beam pipes, and includes reshaped bellows with new designed RF contacts, see Fig. 1.

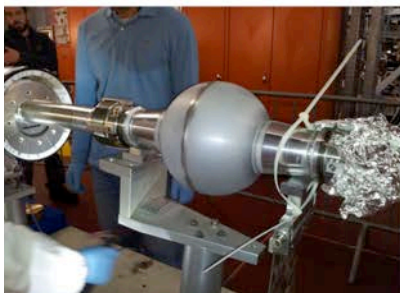

**Figure 1:** IP spherical vacuum chamber.

Replacing the bellows solved the low-β defocusing quadrupole heating problems, recovering working point stability during operations.

Two cooling pipes have been added on the tapers and new semi-cylindrical thin (35 μm) beryllium shields have been placed inside the sphere. Two additional Beam Position Monitors (BPM) have been installed on both sides of the IP, for a more accurate beams overlap and to perform transverse betatron coupling studies.

The IP chamber and the low-β defocusing quadrupoles are suspended at the two sides of the detector. The whole support structure is critical for the stability of the assembly. The design of supports, of the vacuum chambers and equipment, as well as for the magnetic and diagnostic elements have been revised to host the new detector components and the hugely increased number of cables and pipes for gas and coolant, as well as to stand additional weight, see Fig. 2, and improve alignment precision.

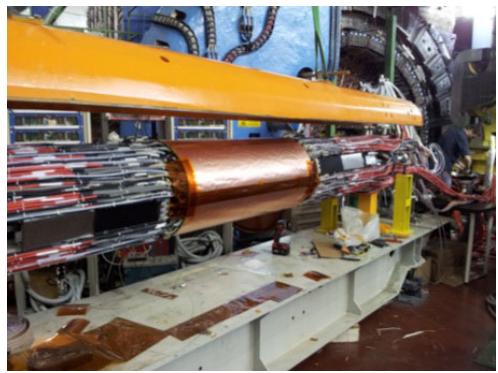

**Figure 2:** The DAΦNE IR with the new detector layers ready to be inserted in KLOE-2.

In particular, a pair of additional carbon fiber composite legs has been designed and superimposed to the existing ones, and some rubber pads previously inserted below the cradle support have been removed, to strengthen the structure and increase its rigidity, see Fig. 3.

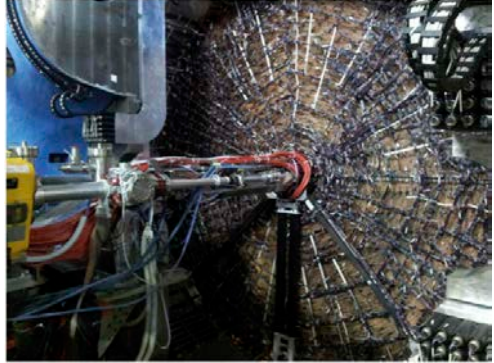

**Figure 3:** Additional carbon fiber composite legs (central black poles) superimposed to the original low-β support structure.

As a result the spectrum of the vertical beam oscillation has changed. The main harmonic has been shifted toward higher frequencies, ~15 Hz, and its amplitude reduced by about a factor three, see Fig. 4 showing the e$^+$ and the e$^-$ oscillation spectra respectively, before and after inserting the new legs.

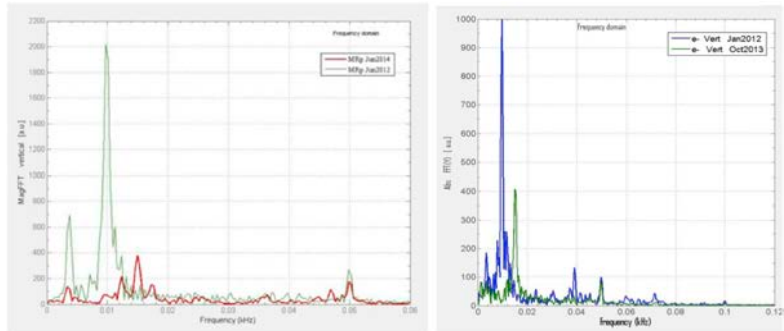

**Figure 4:** Natural e$^+$ beam oscillation spectrum around the nominal orbit as recorded at the BPBPL201 before (green left) and after (red left) revising the low-β support structure. Same analysis for the e$^-$ beam oscillation spectrum as recorded at the BPBEL201 before (blue right) and after (green right) the upgrade.

About the steelwork structure around the IR, some reinforcing plates have been added to the H-shaped girders, including new grounding anchorage with adjustable bolts for the tail of the girders itself.

The screw holes in the jaw of four scrapers, installed two in the e$^-$ and two in the e$^+$ ring at either ends of the IR, have been filled by adding shielding copper extensions, in order to avoid HOM trapping. Since after having detected some anomalous heating events in operations, a visual inspection of the components, revealed clear signs of discharges. Moreover the jaw limit switches have been moved to increase the collimator stroke, thus achieving a more efficient suppression of the background hitting the experimental detector. In total the jaw insertion length has been increased by 2.5 mm, corresponding to a 30% increase with respect to the previous stroke.

Several other developments have been implemented:
- More and better placed CCR holes for alignment have been added.
- Newly designed mechanics (cams and kinematics) now allows a better control of the angular rotation of the low-β focusing quadrupoles from outside the detector.

- Temperature probes have been added on the Interaction Region vacuum chamber.
- Toroidal shields have been added around the IP to reduce the background hitting the detector.
- New Beam Position Monitors have been installed along the rings.

### 2.1.2.2  *Ancillary Plants' Control System Revamping*

The control system of several utility plants serving the accelerator has been renovated. The systems involved have been the Fluids plants (cooling and HVAC, compressed air), the RF plants, the Vacuum plants and the safety system of the magnets over-temperature control subsystem. 15 PLC substations have been replaced. Control system of the Fluids plants has been re-engineered, substituting the whole PLC system.

The control logic of the new system has been modified to take into account the relevant reduction of the AC power demand of the largest magnets (Wigglers, Septa). Replacement of obsolete items drove the change of the PLCs controlling the Vacuum plants (valves and gauges) and the RF plants and interlocks (klystrons, cavities, circulators and loads), but in this case the dismissed equipments have been kept to be used as spare parts for the magnets over temperature control subsystem.

In this refurbishment also the SCADA has been renewed allowing the remote control of all the subsystems to facilitate and speed up faults diagnosis during the current machine operation. The Supervisor was developed with Movicon SCADA and customized upon specific requests.

### 2.1.2.3  *High Pressure Cooling System Optimization*

The water flow rate in the high pressure cooling system serving the wiggler magnets has been reduced by a 33%, by adding a variable frequency drive in the circuit. This allowed definitively avoiding destructive effects induced by cavitation, eroding the copper of the wiggler coils and leading to water leakage. Recovering from this kind of faults required, in general, 2-3 days stop in the operations, and forced to shut down some dipoles and all the wigglers in the main rings in order to access and solder the damaged coils. In addition to the damage in terms of machine uptime, such problems were deteriorating permanently the wiggler coils threatening the magnets long term operativeness.

Cooling system optimization has been possible since in 2010 the poles disposition of the DAΦNE wigglers has been revised [12] achieving, among other, a considerably increase of the magnetic field obtained at a given current. As a consequence the nominal operation current of the wigglers has been remarkably reduced, which allowed, in turn, to relax the cooling system parameters.

### 2.1.2.4  *Control System Upgrade*

The DAΦNE Control System (DCS) has been deeply modified in order to dismiss obsolete components and improve responsiveness and reliability.

In its original design, the system live data resided in a central VME shared memory and the communication channels were based on point-to-point optical links, realized with VME boards. This architecture granted high bandwith and low latency but — on the other hand — was heavily hardware dependent, requiring the use of VME embedded processors for any purpose.

The new design of the DCS hinges on the redirection of the whole data flow to the Ethernet network and the adoption of an *Object Caching* service (*Memcached*) for hosting the live data. This utterly decouples software services from hardware, opening the system to different hardware choices.

Most of the front-end VME boards (serial communication boards, DAQs, ADCs, etc.) have been replaced by network devices, which allowed for the porting of many control programs to remote Linux virtual machines. In particular, the adoption of *serial device servers* instead of serial communication boards, permitted to increase the number of daisy chain lines (RS-422) employed in connecting the magnets' power supplies and consequently to shorten the machine switch (from positrons to electrons and vice-versa).

The DCS upgrade also aimed at replacing the original distributed front-end VME processors (forty-five 68030 custom boards, running MacOS 7) with Intel boards, running Linux, see Fig. 5. At the present, 19 Virtual Machines and 7 new Linux VME processors host 70 control processes and only 10 of the former VME processors are left.

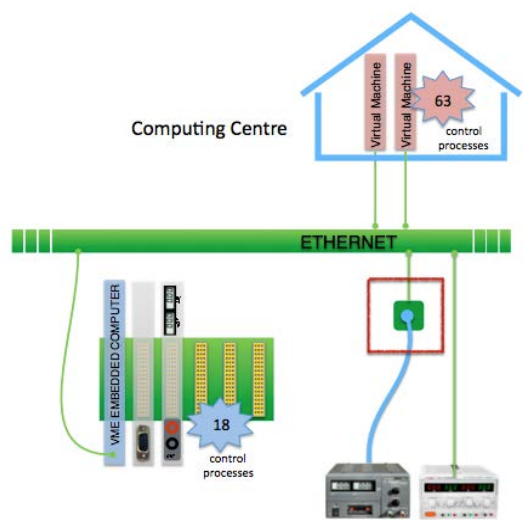

**Figure 5:** DAΦNE control system new layout.

New Linux servers have been setup for the core services (NFS, DHCP, diskless boot, MySQL, memcached) and for the SunRay™ thin-clients employed as consoles. The system hosting the virtual machines has been set-up with Red Hat Cluster Suite and XEN 3.2014.

The Network Uplink with the Computing Centre has been enhanced (10 G) in order to improve the command/data flow among virtual machines and front-end devices.

The upgrade has also concerned both the *hardware* and the *software* of many front-end systems, in order to take advantage of the new DCS structure. The DAΦNE subsystems that have gone through major changes are: RF Slow Control, Main Ring Scrapers, Programmable Delays, Power Supplies, Main Rings and Damping Ring Kickers, Beam Charge Monitors, Spectrometer, Vacuometers, Vacuum Pumps and Clearing Electrodes.

After the upgrade, the DCS proved to be performable and reliable and its overall uptime - in real operating condition - significantly increased.

### 2.1.2.5 *Cryogenic Plant*

The cryogenic plant, serving the superconducting solenoid of the experimental apparatus and the four anti-solenoids installed on the collider rings, has been completely overhauled and some specific parts have been mended or replaced. Some o-rings sealing in the helium transfer lines have been replaced with soldered connections. Two partially damaged Joule-Thomson needle valves have been reworked. A remotely controlled pneumatic valve has been added in the liquid nitrogen line. PT100 thermometers were installed in the nitrogen line of the anti-solenoids transfer lines, close to the gas flow controllers. The obsolete remote PC for the plant control has been replaced as well as the operator interface panel. The listed activities were aimed at preventing accidental freezing at the controller level, and at ensuring remote procedures for refilling the anti-solenoids. Nevertheless, other long shutdown periods have been necessary, afterwards, to recover from an oil contamination at the level of the cold-box, and to undertake an extraordinary maintenance of the compressor, which is working since 750000 hours, well beyond its expected working lifetime.

### 2.1.2.6 *Linac*

All the LINAC components have been overhauled paying special attention to the four RF power plants. In this context several exhausted components such as filter capacitors, thyratrons and high power pulse discrete elements have been replaced, and a new designed RF driver system has been installed aiming at achieving a better stability in terms of delivered power.

New vacuum pumps and ancillaries have been added on the four main waveguides downstream the SLEDs, in order to reduce discharge occurrences.

Concerning the RF-vacuum devices in the LINAC accelerating sections, the residual pressure considerably improved by replacing all the RF loads. The vacuum safety system gating valves and some in-vacuum diagnostic elements such as flags and BPMs have been also replaced.

All the ceramic windows, placed downstream the klystron ones to decouple the LINAC vacuum, were almost at the end of their operating-life and have been preventively substituted.

As a special case, the RF power plant 'D', driving the last four accelerating sections, required an extraordinary mending effort, even beyond the 2013 shutdown. Several parts had to be replaced such as: the klystron, the waveguide elbow interfacing the klystron, the SLED and many ancillary components.

Some bugs in the Helmholtz Coil power supplies have been detected and fixed. The PLC control system has been upgraded, and its parts underwent an accurate revision involving: water ducts, flux-meters and water pumping system, leading to replacement of many components. The LINAC control system has been revised and upgraded in order to be compliant with the renewed Ethernet architecture (new routers and VLAN relying on fiber connections) and to profit from new network features. In this context a new control application, based on dedicated multiplexed DAQ, has been designed and implemented for the 14 LINAC BPMs.

### 2.1.2.7 *Other Consolidation Activities*

Many other remarkable activities have been done.

The magnetic field of the *IR defocusing quadrupoles* has been measured detecting discrepancies of the order of few % only with respect to *ab initio* characterization.

The 32 power supplies powering a family of corrector magnets have been replaced with updated devices.

The HV power supplies polarising the *e-cloud clearing electrodes* have been substituted with devices providing twice the original voltage and having negative polarity. This allows achieving a complete neutralization of the *e-cloud* generated by a positron current of the order of ~1. A, and reducing the generator delivered current.

More robust *feedthroughs* have replaced the ones originally used for the electrodes installed inside the wiggler magnets of the e$^+$ ring.

Two *new vacuum chambers* have been built and installed near the injection sections of both rings. Each new beam pipe is carrying eight button BPMs to be used for orbit measurements and as feedback pickups.

One of the two klystrons, stored in the DAΦNE hall to serve, in the case, as a spare part for the RF plants of the DAΦNE main rings, lost vacuum insulation.

A thorough analysis pointed out that purchasing a spare tube, same as the damaged one, was considered unworthy since costs and time required for this acquisition resulted to be simply unaffordable. The only viable way to recover the broken tube consisted in finding and repairing the vacuum leakage in house. The repaired tube has been installed in the e$^-$ ring power plant in winter 2015 to replace another broken unit, and after a brief period of conditioning, it has reached the nominal performance required for running the machine and it is presently in operation.

Concerning the bunch-by-bunch feedback systems, a *new horizontal kicker* with a doubled stripline length has replaced the original one on the electron ring, providing larger shunt impedance at the low frequencies typical of the unstable modes. This allows doubling the feedback damping rate for the same setup (gain, power amplifier, etc.). A dedicated virtual LAN for all feedback units provides a faster real-time data processing. Hardware has been upgraded and Linux software updated to be compliant with the most recent netware and software releases.

### 2.1.3 DAΦNE Main Rings Tuning

Operations, in general, received powerful impulse from the consolidation activities.

The work done on the IR mechanical structure, for instance, had a huge impact on many main rings crucial issues such as: impedance budget, optics and beam dynamics. Similarly the numerous mending actions involving almost all the DAΦNE subsystems have been essential in restoring a good uptime, a fundamental prerequisite to undertake reliable measurements and for fine tuning. In a word, without these upgrades configuring machine for collisions, tuning luminosity and, as a matter of fact, testing the Crab-Waist collision scheme with a large detector would have not been possible.

#### 2.1.3.1 *Colliding Rings Optics*

The IR layout [8] implementing Crab-Waist collisions for the KLOE-2 detector includes a low-β section based on permanent magnet quadrupole doublets. The quadrupoles are made of SmCo alloy: the first one from the IP has a gradient of 29.2 T/m, the second 12.6 T/m. The first is horizontally defocusing and is shared by the two beams; due to the off-axis beam trajectory, it increases the horizontal crossing angle from ~25 to ~ 50 mrad. In addition two anti-solenoids are installed symmetrically with respect to the IP in each ring. The new ring optics account for all these features, and, at the same time, assure suitable betatron oscillation amplitude at the *Crab-Waist* Sextupoles (*CW*-Sextupoles), and proper phase advance [4] between these magnets and the IP.

### 2.1.3.2 *Transverse Betatron Coupling Correction*

The permanent magnet focusing quadrupoles of the low-β are rotated around their longitudinal axes, as well as the three electromagnetic quadrupoles installed on each one of the four IR branches. These rotations, together with the four anti-solenoids, provide an efficient compensation mechanism for the coupling due to the solenoid of the experimental apparatus [8]. Moreover the rotations of the low-β focusing quadrupoles, independent for the two rings, are used for transverse betatron coupling fine tuning. The procedure relies on transverse beam size measurements as evaluated by a calibrated synchrotron light monitor, and he response matrix measured by varying corrector magnets. Presently a very good coupling correction has been achieved for the $e^+$ beam, $\kappa \sim 0.4\%$, with all the skew quadrupoles off, while a further optimization is needed for the $e^-$ beam. Nevertheless by tuning the skew quadrupoles a transverse betatron coupling in the range $0.2\% \div 0.3\%$ can be achieved in both rings [13].

### 2.1.3.3 *Main Rings Optics and Working Point Studies*

Main rings optics studies profited a lot from working point stabilization, which has been recovered by replacing broken bellows, close to the IP. This was causing anomalous heating (up to $50 \div 60\ ^0C$) in the low-β defocusing quadrupole downstream the $e^-$ beam, and in turn, a random significant oscillation of the tunes in the main rings, especially the vertical one. Fixing this aspect allowed to undertake reliable and systematic machine measurements, aimed at methodical optimization of main rings lattice, working point and collisions.

Working point simulations at DAΦNE are performed by using the *Lifetrack* simulation code. *Lifetrack* is a weak-strong particle tracking code, developed for simulating equilibrium density distributions in lepton colliders [14]. The fully symplectic 6D treatment of beam-beam interaction allows to simulate configurations with very large crossing angle, and crabbing of weak and strong bunches either. Main code features, in the equilibrium distribution case, allows computing: 3-D density of the weak beam, specific luminosity, beam lifetime, dynamical aperture and Frequency Map Analysis [15]. Moreover, latest code developments implement a detailed machine lattice model, via element-by-element tracking in thin lens approximation. The optics description for both weak and strong beam can be imported directly from MAD-X [16] model files. Such approach enables to properly treat element misalignments and related orbit distortions, chromatic aberrations, lattice nonlinearities, betatron and synchrobetatron coupling, with the same formalism used for ordinary optics simulations.

The working points initially devised for the DAΦNE main rings were: $\nu_x^- = 5.098$, $\nu_y^- = 5.164$ and $\nu_x^+ = 5.1023$, $\nu_y^+ = 5.139$, which, according to simulations, provide good luminosity and lay in a rather large stable area. In fact, with this optics a luminosity of $1.88\ 10^{32}$ has been achieved.

This configuration, in its early stage, when the highest achievable luminosity was at last $1.5 \cdot 10^{32}\ cm^{-2}s^{-1}$, has been used for a test run using 10 colliding bunches. The study, see Fig. 6, lasted for about 10 hours, during which a peak luminosity of $L = 2.55 \cdot 10^{31}\ cm^{-2}s^{-1}$ has been repeatedly measured with currents of the order of $I^- \sim 0.096$ A, and $I^+ \sim 0.126$ A, without any particular optimization effort. Since single bunch current is comparable with the one used during nominal high current, high number of bunches operations, this experiment provides a nice environment to disentangle the main beam-beam contribution to the maximum achievable luminosity, from the components due to collective effects dominating the DAΦNE beam dynamics. In the specific case the measurement indicates that a luminosity of the order of $L = 2.5 \cdot 10^{32}$ cm$^{-2}$s$^{-1}$ is in principle achievable.

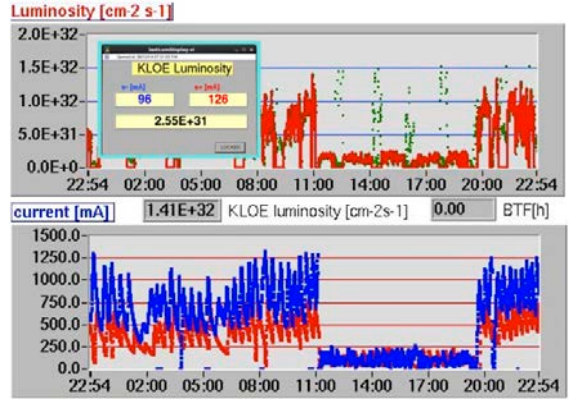

**Figure 6:** 10 bunches luminosity test.

In spite of the positive achievements, still some limitation persisted, affecting mainly the e$^-$ beam in terms of injection efficiency, beam lifetime and e$^-$ beam induced background. New simulation outlined that moving the e$^-$ ring tunes to new values ($\nu_x^- = 5.13$, $\nu_y^- = 5.17$) would have led to improve dynamical aperture by 2-3σ, suppressing at the same time the growth of the bunch vertical tail and achieving a moderate increase in terms of specific luminosity too [17]. Changing the working point required to compute and implement a new optics, which, in turn, imposed to optimize the transverse feedback systems. The new configuration led to very positive results. The contribution of the e$^-$ beam to the total machine background was reduced by 30% and 20% [18], as can be seen from the left graph in Fig. 7, which shows a comparison between machine background hitting the KLOE-2 calorimeter, in the region around the exit of the electron beam, as a function of the instantaneous luminosity. The KLOE-2 trigger rate also profited from the new electron ring optics, as it is evident from the right plot in Fig. 7. The observed reduction of the trigger rate had a twofold positive effect. First, the data throughput decreased, since the contribution of the machine background hitting the KLOE-2 sub-detectors to the event size is smaller, and because the events acquisition rate is smaller. Second, the dead time induced by the activation of the KLOE-2 trigger is smaller, allowing for a more efficient data taking.

In addition to the background reduction, after short period spent on collider fine-tuning, the best peak luminosity, $L_{peak} = 2.0 \cdot 10^{32}$ cm$^{-2}$ s$^{-1}$, and the best daily integrated luminosity, 12.5 pb$^{-1}$, ever measured with the KLOE-2 detector have been attained in a stable and reproducible way. Repeating the 10 colliding bunches test would be very instructive, but for now the data taking schedule imposes drastic limits on the time available for machine studies.

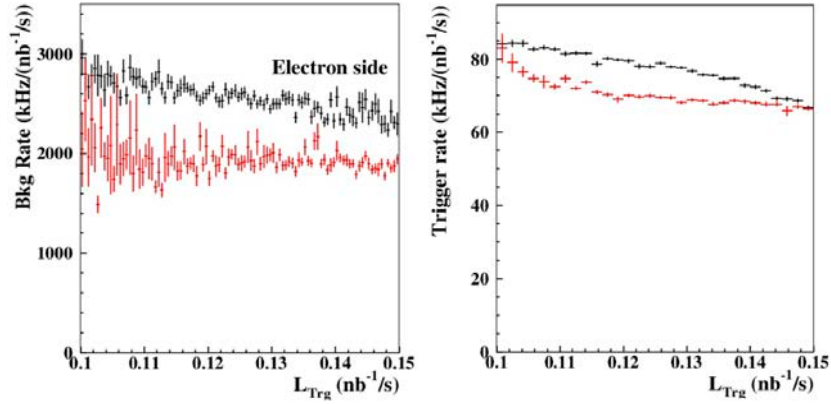

**Figure 7:** Comparison between machine background hitting the KLOE-2 calorimeter, in the region around the e⁻ beam exit, as a function of instantaneous luminosity (left), and KLOE-2 trigger level (right) for previous (black dots) and current (red squares) e⁻ machine working point. KLOE-2 trigger provides the instantaneous luminosity measurement. The reduction of background rate as a function of luminosity for the previous DAΦNE optics is mainly due to dynamical effect induced by the high current (~1A) needed to reach high luminosity. The reduction in the KLOE-2 trigger rate observed in the right panel has a similar behaviour as the background rate.

### 2.1.4 Beam Dynamics

Machine operation at high current strongly depends on vacuum conditions. Since the main rings beam pipe has been opened, repeatedly, in several sections a quite long time has been spent to recover a reasonable dynamic vacuum level.

Highest currents stored, so far, are 1.7 A and 1.2 A, for e⁻ and e⁺ beam, respectively. These currents are the highest ever achieved after installing the new IR for the KLOE-2 detector, based on the *Crab-Waist* collision scheme.

The three independent bunch-by-bunch feedback systems [19] installed on each DAΦNE ring are continuously working being essential for high current multi-bunch operations. The e⁺ vertical feedback is now using a new ultra-low noise front-end module, designed in collaboration with the SuperKEKB feedback team, aimed at reducing the noise contribution to the transverse vertical beam size in collision.

Presently beam dynamics in the e⁺ ring is clearly dominated by the e-cloud induced instabilities, whose effects are suppressed by means of powerful bunch-by-bunch transverse feedback systems [20], by solenoids wound all around the straight sections and by on purpose designed electrodes [21] installed inside dipole and wiggler vacuum chambers. The electrodes have been already checked in 2012, during the KLOE preliminary run. Several measurements and tests demonstrated their effectiveness in thwarting the e-cloud effects [22]. These first studies have all been done by biasing the striplines with a positive voltage in the range 0÷250 V. However simulations indicate that a factor two higher voltage is required to completely neutralize the *e-cloud* density due to a e⁺ current of the order of 1 A. For this reason, during the 2013 shutdown, the electrode power supplies have been replaced with devices providing a maximum negative voltage of 500 V. The change of polarity was intended to limit the current delivered by the power supplies. The new setup has been tested storing a ~ 700 mA current in 90 bunches spaced by 2.7 ns, and measuring the horizontal and vertical tune spread along the batch with the electrodes on and off. Results show a clear reduction of the tune spread in both planes, but especially in the horizontal one [23]. It's worth mentioning that presently three out of the four electrodes installed in the wiggler magnets have been short-circuited since they were not working properly. Moreover a random vertical oscillation of the e+ beam orbit, affecting the collider fine tuning and data delivery, has been correlated with the electrodes operation. Orbit perturbations with amplitude in the range ±0.5 mm have been already observed before the 2013 shutdown, albeit with lower occurrence, and have been ascribed to the electrodes for two experimental reasons. First, oscillation phase can be decomposed in terms of the phases characterizing the orbit variations measured switching on the electrodes one by one. Second, the effect has been cured, after achieving a reasonable dynamic vacuum level, by tuning the electrode working voltage in order to limit the current delivered by the power supply. However, the mechanism causing this effect has not yet been completely understood.

Another positive result in mitigating the detrimental effects induced by the e-cloud has been obtained lengthening the bunch by reducing the voltage of the RF cavity of the $e^+$ ring. Fig. 8 presents the behaviour of the pressure rise with the stored current, measured by two vacuum gauges installed on different arcs, as a function of the RF cavity voltage.

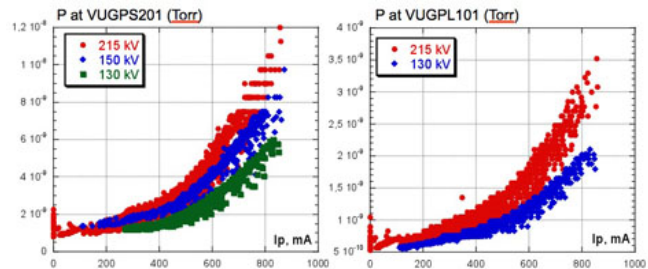

**Figure 8:** Pressure rise versus stored $e^+$ current in two different arcs of the $e^+$ ring as a function of the RF cavity voltage.

The e- beam exhibits a microwave instability threshold (TMCI), appearing above a current of the order of ~10 mA per bunch, resulting in a widening of the transverse beam sizes. Such effect is quite moderate in single beam operation and becomes more harmful in collision due to the beam-beam interaction. The instability might be limited by implementing an optics providing a higher value of the momentum compaction $a_c$.

In general beam dynamics also profited from upgrading collimators and replacing the bellows installed in the IR close to the low-β section. In fact some of them were seriously damaged and were causing random discharges.

### 2.1.5  *Crab-Waist* Collisions

A detailed comparison of the beam parameters corresponding to the record luminosities achieved in some topical stages of the DAΦNE activity is presented in Tab. 1.

Maximum instantaneous luminosity is now a ~33% higher with respect to the past KLOE run, regardless it has been obtained by colliding beams having lower currents and lower number of bunches. This improvement is consistent with the maximum daily integrated luminosity, which is now ~28% higher with respect to 2005.

Despite these positive results, instantaneous luminosity is still a factor 2 lower than the peak value measured during the Crab-Waist test run. Anyhow daily integrated luminosity differs, in defect, from the best attained in 2009 by a ~20% only.

**Table 1:** Beam currents and number of bunches used for collisions: during the test run of the new Crab-Waist collision scheme, while giving data to a detector without solenoidal field, during the KLOE run in 2005 and in the present configuration which integrates the Crab-Waist collision scheme with the KLOE-2 detector having high perturbing solenoidal field.

| Parameter | Crab-Waist test run (2009) | KLOE run (2005) | KLOE-2 Crab-Waist (2015) |
|---|---|---|---|
| $L_{peak}$ [cm$^{-2}$s$^{-1}$] | $4.53 \cdot 10^{32}$ | $1.5 \cdot 10^{32}$ | $2.0 \cdot 10^{32}$ |
| $I^-$ [A] | 1.52 | 1.4 | 1.03 |
| $I^+$ [A] | 1.0 | 1.2 | 1.03 |
| $N_b$ | 105 | 111 | 103 |
| $L_{/day}$ [pb$^{-1}$] | 15 | 9.8 | 12.5 |

Several solid arguments, based on experimental and theoretical considerations, indicate that the higher instantaneous luminosity measured with the KLOE-2 detector is due to the beneficial effects introduced by the *Crab-Waist* collision scheme.

A comprehensive numerical study has been performed in order to investigate the effectiveness of the crab waist collision scheme in presence of the strong solenoidal fields introduced by the KLOE-2 detector [17]. Beam-beam simulations have been carried out taking into account the real DAΦNE nonlinear lattice including the detector solenoidal fields, IR quadrupole rotations, compensating anti-solenoids etc. The numerical results clearly show that, in DAΦNE, the detector solenoid does not determine any relevant reduction in terms of crab waist collision scheme effectiveness.

Indeed, *CW*-Sextupoles have been used since the beginning of the collider commissioning with KLOE-2, then their strength has been gradually increased along with the nonlinear beam dynamics optimization.

Experimental test [13], done switching off the *CW*-Sextupoles, lead to achieve a luminosity slightly in excess of $10^{32}$ cm$^{-2}$s$^{-1}$, and outlined harmful effects such as: beam lifetime reduction while injecting the opposite beam, transverse beam blow-up at high current and unprecedented background level, both in coasting and injection regime. All these phenomena are perfectly consistent with the lack of a compensation mechanism for the synchro-betatron resonances affecting collisions in large Piwinsky angle regime.

The beneficial effect of the *CW*-Sextupoles can be clearly seen by looking at the luminosity, lifetime and the detector background by varying the *CW*-Sextupoles strength see Fig. 9.

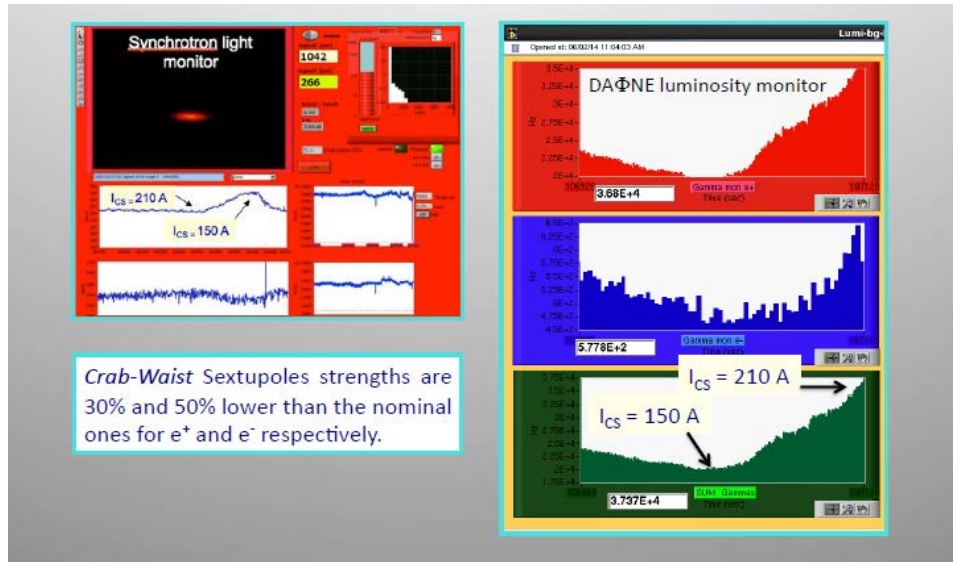

**Figure 9:** Decreasing the *CW*-Sextupoles strengths the vertical beam sizes, as measured at the synchrotron light monitor, increase (left panel); at the same time the luminosity grows, as reported by the DAΦNE fast luminosity monitor (right panel upper and central frame) and by the KLOE detector luminosity monitor (right panel lower frame).

However the full potential of the crab waist collision scheme has not yet been exploited due to several limiting factors already discussed in the previous paragraphs.

In fact future programs foresee to improve DAΦNE performances by:

- improving *CW*-Sextupoles alignment on the beam orbit and optimizing their strengths
- refining transverse betatron coupling correction
- pushing the microwave instability threshold toward higher single bunch current value by means of new optics configuration having higher $\alpha_c$ and higher chromaticity
- vacuum conditioning and beam scrubbing to diminish the e-cloud impact on e$^+$ beam dynamics
- further feedback noise reduction
- tuning the interplay between RF 0-mode feedback and longitudinal feedback.

### 2.1.6   DAΦNE Data Delivering

On mid November 2014 a plan has been agreed with the KLOE-2 collaboration team in order to start a preliminary data-taking campaign. The most relevant point of the plan stated the collider had to deliver 1 fb$^{-1}$ in 8 months long continuous operations.

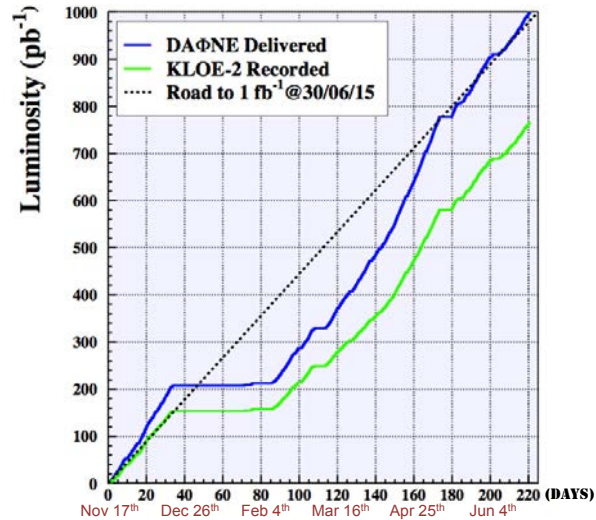

**Figure 10:** Delivered (blue line) and acquired (green line) luminosity presented together with the scheduled plan (black dotted line).

Data taking results are summarized in Fig. 10. In the first month of operations the integrated luminosity growth rate was well above the predefined guideline. After that, the 23 days long stop, scheduled for Winter holiday prolonged to mid February due to delays in completing some activities, and to several faults occurring in sequence and having different relevance. Main faults involved the cooling system of the KLOE magnet power supply and the klystron of the RF plant serving the electron ring. Although the first problem required long time, a lot of measurements and chemical wash of the circuit to be fixed, it did not pose any concern on the collider program feasibility. The latter, on the contrary, was quite threatening, since it required the installation and testing of the klystron previously mended in house and stored in the DAΦNE hall. The whole operation, including: faulty klystron removal, spare device installation and conditioning, took about 10 days. Whereupon a stable electron beam with a current Γ~ 1.6 A has been stored and used for collisions. Tests performed on the broken klystron pointed out a leakage in the insulation vacuum, very much similar to the one already repaired, which was fixed as well.

In the following three months DAΦNE had long, stable operations, in which achieved its best performances in terms of instantaneous and integrated luminosity.

Collider optimization, progressive vacuum melioration, luminosity fine tuning, higher number of colliding bunches and improved control over e-cloud induced effects led instantaneous luminosity to reach the value of $1.88 \ 10^{32}$ cm$^{-2}$ s$^{-1}$ by mid March 2015. Thence, after adopting a new more suitable working point in the electron ring, according theoretical simulation, a reproducible peak luminosity of the order of $2.0 \ 10^{32}$ cm$^{-2}$ s$^{-1}$ has been achieved by the end of April.

Stability of the collider setup is confirmed by record results concerning long term integrated luminosity. In fact, 71 bp$^{-1}$, see Fig. 11, have been delivered in a week of operation.

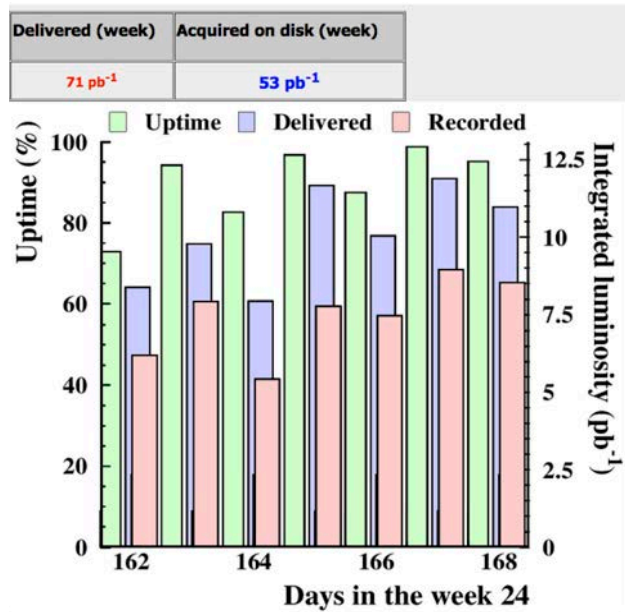

**Figure 11:** Best weekly delivered (violet bin) and integrated (red bin) luminosity, presented together with the machine uptime (green bin) defined as the percent fraction of the day the collider has been delivering luminosity, suitable for acquisition, in excess of ~ 0.1 $10^{32}$ cm$^{-2}$ s$^{-1}$.

This promising result when accounted together with the 251 pb$^{-1}$ delivered in 30 consecutive days, see Fig. 12, clearly indicates that the KLOE-2 run, aimed at collecting about 5 fb$^{-1}$ total integrated luminosity, is feasible and can be completed in a time lapse of the order of 2-3 years.

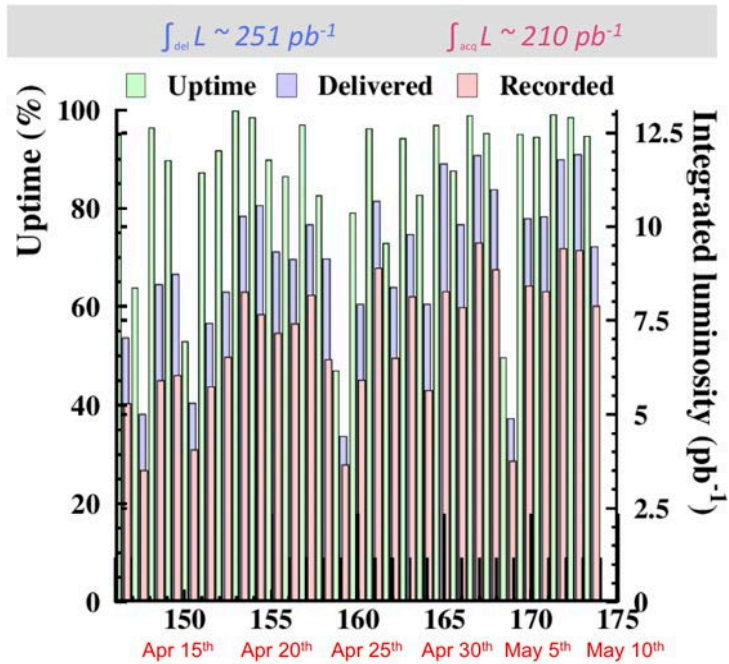

**Figure 12:** Best monthly delivered (violet bin) and integrated (red bin) luminosity, presented together with the machine uptime (green bin).

Furthermore long term integrated luminosity can still be improved, even without changing currents and number of colliding bunches, as suggested by the best hourly integrated luminosity measured averaging over two hours, which is L$_{\int 1hour}$ ~ 0.54 pb$^{-1}$, see Fig. 13, regardless some minor faults occurring during the run and some inefficiency in the injection process.

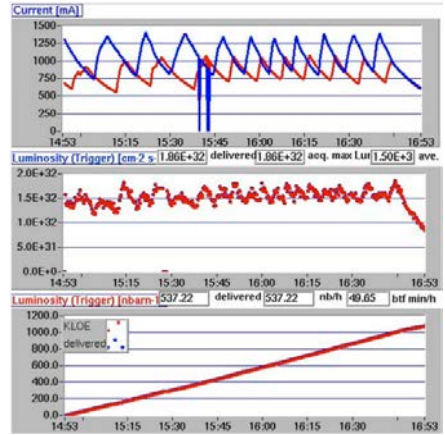

**Figure 13:** Best hourly integrated luminosity.

Collider uptime, presented in Fig. 11 and Fig. 12, is defined as the percent fraction of the day in which the collider has been delivering luminosity, suitable for acquisition, in excess of ~ 0.1 10$^{32}$ cm$^{-2}$ s$^{-1}$. A quite strict definition indeed. Nevertheless average uptime, in the interval presented, is always well above 80%, a value representing the upper limit usually achieved in particle accelerators like DAΦNE.

Another issue of primary importance, second only to the integrated luminosity rate, is the background hitting the experimental apparatus. Background produced by the beams colliding at DAΦNE is essentially due to the Touschek effect. It has a very high impact on all subsystems involved in the data acquisition: DAQ boards, KLOE-2 networks, data buffering, mid-term (disk) and long-term (tape) support consumption. Although the present background rate is compatible with an efficient detector data taking, it is about a factor two higher with respect to the old KLOE run (2005). Special concern was given by the component due to the e- beam, which has been considerably reduced varying optics in the e$^{-}$ ring.

In the last two months integrated luminosity growth rate slowed a little bit down due to periodic maintenance, safety checks and the exceptionally high atmospheric temperatures forcing several subsystems to work in critical condition.

In summary, 1.0 fb$^{-1}$ has been delivered to the experiment in approximately 200 days of activity according the scheduled plan. Moreover integrated luminosity delivering rate, background level and collider uptime are compatible with an efficient data taking of the KLOE-2 detector.

### 2.1.7 Conclusion

The DAΦNE collider has recently achieved very positive results: instantaneous luminosity and integrated luminosity rate are now the highest ever measured in operations with an experimental apparatus including high field detector solenoid, confirming the *Crab-Waist* collision scheme effectiveness in achieving high luminosity even in presence of a large detector.

Machine uptime profited from the several consolidation activities implemented, and presently can assure a long term data taking to the detector.

Beside the present promising results, several limiting factors have been outlined and understood, and still many parameters can be ameliorated to improve the collider performances.

The KLOE-2 collaboration has started its data acquisition campaign, and the collider has already delivered $1\,\mathrm{fb}^{-1}$ according the scheduled plan.

## 2.1.8 **References**

1. G. Vignola et al., "Status Report on DAΦNE", Frascati Phys. Ser. 4:19-30, 1996; C. Milardi et al., "Status of DAΦNE", Frascati Phys. Ser. 16:75-84, 1999.
2. C. Milardi et al., "DAΦNE Operation with the FINUDA Experiment", EPAC04, pp. 233-235, arXiv:physics/0408073.
3. A. Gallo et al., "DAΦNE Status Report", EPAC06, Conf.Proc. C060626 (2006) 604-606.
4. P. Raimondi et al., "Beam-Beam Issues For Colliding Schemes with Large Piwinski Angle and Crabbed Waist", arXiv:physics/0702033.
5. C. Milardi et al., "Present Status of the DAΦNE Upgrade and Perspectives", Int.J.Mod.Phys.A24:360-368, 2009.
6. M. Zobov et al., "Test of Crab-Waist Collisions at DAΦNE Phi Factory", Phys. Rev. Lett. 104, 174801 (2010).
7. C. Milardi et al, "DAΦNE Developments for the KLOE-2 Experimental Run", IPAC10, p. 1527, Conf.Proc. C100523 (2010) TUPEB006.
8. C. Milardi *et al.*, "High Luminosity Interaction Region Design for Collisions Inside High Field Detector Solenoid", 2012 *JINST* 7 T03002.
9. C. Milardi et al, "DAΦNE Tune-Up for the KLOE-2 Experiment", IPAC11, Conf.Proc. C110904 (2011) 3688-3690.
10. KLOE-2 Collab., LNF - 10/14(P) April 13, 2010.
11. C. Milardi et al, "DAΦNE General Consolidation and Upgrade", IPAC14, p. 3760, IPAC-2014-THPRI002.
12. S. Bettoni et al., "Multipoles Minimization in the DAΦNE Wigglers", IPAC10, Conf.Proc. C100523 (2010) THPE065.
13. C. Milardi et al, "DAΦNE Operation with the Upgraded KLOE-2 Detector, IPAC14, p. 1883, IPAC-2014-WEOCA03.
14. D. Shatilov, "Beam-Beam Simulations at Large Amplitudes and Lifetime Determination", Particle Accelerators 52:65-93, (1996).
15. D. Shatilov et al., "Application of Frequency Map Analysis to Beam-Beam Effects Study in Crab Waist Collision Scheme", Phys. Rev. ST Accel. Beams 14, 014001 (2011).
16. **http://cern.ch/madx**
17. M. Zobov et al., "Simulation of Crab Waist Collisions In DAΦNE with KLOE-2 Interaction Region", IPAC15, p. 229, arXiv:1506.07559.
18. A. De Santis, https://agenda.infn.it/conferenceDisplay.py?confId=9613.
19. A. Drago et al., "Recent Observation on a Horizontal Instability in the DAΦNE positron ring", PAC05, p. 1841, SLAC-PUB-11654.
20. A. Drago et al., "DAΦNE Bunch by Bunch Feedback Upgrade as SuperB Design Test", IPAC11, Conf.Proc. C110904 (2011) 490-492.
21. D. Alesini et al., "DAΦNE Operation with Electron-Cloud-Clearing Electrodes", Phys. Rev. Lett. 110, 124801 (2013).

22. D. Alesini et al., "Design and Test of the Clearing Electrodes for e-Cloud Mitigation in the e+ DAΦNE Ring", IPAC10, p. 1515, Conf.Proc. C100523 (2010) TUPEB002.

23. A. Drago, http://indico.cern.ch/event/306551/.